# A Bayesian Evidence Synthesis Approach to Estimate Disease Prevalence in Hard-To-Reach Populations: Hepatitis C in New York City

Sarah Tan[1], Susanna Makela[2], Daliah Heller[3], Kevin Konty[4], Sharon Balter[5], Tian Zheng[2], James H. Stark[6]

[1] Department of Statistics, Cornell University; ht395@cornell.edu
[2] Department of Statistics, Columbia University
[3] Graduate School of Public Health and Health Policy, City University of New York
[4] New York City Department of Health and Mental Hygiene
[5] Los Angeles County Department of Public Health
[6] Worldwide Safety and Regulatory, Pfizer, Inc.

## Abstract

Existing methods to estimate the prevalence of chronic hepatitis C (HCV) in New York City (NYC) are limited in scope and fail to assess hard-to-reach subpopulations with highest risk such as injecting drug users (IDUs). To address these limitations, we employ a Bayesian multi-parameter evidence synthesis model to systematically combine multiple sources of data, account for bias in certain data sources, and provide unbiased HCV prevalence estimates with associated uncertainty. Our approach improves on previous estimates by explicitly accounting for injecting drug use and including data from high-risk subpopulations such as the incarcerated, and is more inclusive, utilizing ten NYC data sources. In addition, we derive two new equations to allow age at first injecting drug use data for former and current IDUs to be incorporated into the Bayesian evidence synthesis, a first for this type of model. Our estimated overall HCV prevalence as of 2012 among NYC adults aged 20-59 years is 2.78% (95% CI 2.61-2.94%), which represents between 124,900 and 140,000 chronic HCV cases. These estimates suggest that HCV prevalence in NYC is higher than previously indicated from household surveys (2.2%) and the surveillance system (2.37%), and that HCV transmission is increasing among young injecting adults in NYC. An ancillary benefit from our results is an estimate of current IDUs aged 20-59 in NYC: 0.58% or 27,600 individuals.

## 1    Introduction

Estimating the prevalence of chronic hepatitis C (HCV) in the general population presents numerous challenges. Hepatitis C's asymptomatic nature means many individuals are only tested years after infection [Maheshwari 2008]. Mechanisms used to estimate disease prevalence, including surveillance and surveys, whether serological or self-reported, often under-represent or fail to capture subpopulations at greatest risk for HCV such as former and current injecting drug users (IDUs), the incarcerated, and the homeless [Balter 2014]. This sampling bias results in estimates that are not representative of the general population, necessitating adjustments. However, the quality of these adjustments and uncertainty of resulting estimates are typically unknown [Bornschlegel 2009].

A different approach involves implementing alternative sampling strategies, such as sampling IDUs in drug treatment programs [Jordan 2015] and individuals in correctional facilities, to better reach under-represented subpopulations. These subpopulation estimates are then generalized to the general population. However, this approach invites additional challenges, as some subpopulation sizes – e.g. the number of injecting drug users - are not easily measured, and estimates, when available, are subject to considerable uncertainty [Hickman 2013].

Yet another approach is predictive models based on nationwide data. However, these models fail to capture the inherent spatial-temporal aspects of disease transmission [Wasley 2000]; Bornschlegel et al. suggested important differences at the local scale [Bornschlegel 2009], and multiple projects have concluded that HCV prevalence is higher in NYC than nationally [NYCDOHMH 2013]. Finally, back calculation, in which past HCV prevalence is re-constructed from data on reported deaths from HCV and knowledge of hepatitis C disease progression, has the limitation of greater uncertainty around the estimation of recent prevalence given the long time between HCV infection and death [Sweeting 2007], and is also subject to biases such as the under-reporting of deaths from HCV [Trubnikov 2011].

This paper addresses the limitations of existing methods in two important ways. First, we use Bayesian evidence synthesis to combine multiple sources of data in a systematic manner, account for bias in certain data sources, and provide an estimate of uncertainty associated with the estimates. Specifically, we use the Bayesian evidence synthesis model developed by Sweeting et al. [Sweeting 2008; Sweeting 2009; De Angelis 2009; Harris 2012] to estimate chronic HCV prevalence in the NYC population by IDU risk group and age. This approach results in an improved estimate of HCV prevalence in NYC that is also more inclusive, utilizing ten data sources the New York City Department of Health has on IDU and HCV prevalence, even biased ones not included in previous approaches. Second, we derive two new equations to allow age at first injecting drug use data for former and current IDUs to be incorporated into the Bayesian evidence synthesis model, a first for this type of model.

## 2 Method

We start by defining an estimator for overall HCV prevalence then describe the motivations for using Bayesian evidence synthesis to estimate it.

### 2.1 A Stratified Estimator for HCV Prevalence

The primary mode of HCV transmission in NYC is injecting drug use [NYCDOHMH 2013]. One way to account for this by stratifying the NYC population into three IDU risk groups: $g$ = current, ex, or non-IDU[1]. Since both injecting drug use and HCV prevalence vary by age, the NYC population is further stratified into four age groups: $a$ = 20-29, 30-39, 40-49, or 50-59 years. This two-level stratification results in twelve groups (three IDU risk groups by four age groups). Overall HCV prevalence $\pi$ can now be estimated as a weighted average of two components: HCV prevalence among type $g$ IDUs aged $a$ $\pi_{g,a}$, and the proportion of type $g$ IDUs among individuals aged $a$ $\rho_{g,a}$:

$$\pi = \frac{\sum_a (N_a \sum_g \rho_{g,a} \pi_{g,a})}{\sum_a N_a} \quad (1)$$

where weights $N_a$, the population size of age group $a$ in NYC, are used instead of population sizes of IDU risk groups because information on the former is readily available from the Census [Census 2010] whereas the latter has to be estimated.

---

[1] Current IDUs are individuals who injected at least once in the past year; ex-IDUs injected at least once in their lifetime but not in the past year; non-IDUs have never injected.

## 2.2 Bayesian Evidence Synthesis

An evidence synthesis approach is suitable for this problem because while no single representative data source is available for overall HCV prevalence, partial information is available from different data sources. These data sources cover different NYC subpopulations and are subject to different biases. Some provide unbiased information (e.g. $\rho_{non,a}$ from a household survey); others provide biased information (e.g. $\pi_{ex,a}$ from a community health clinic for chronic diseases). A third possibility is where a data source is biased for one quantity but unbiased for another (e.g. a drug treatment clinic with current injecting drug use not representative of the population $\rho_{cur,a}$, but unbiased for HCV prevalence in IDUs $\pi_{cur,a}$). The last possibility is where information is mixed across IDU risk groups or age groups (e.g. surveys that do not differentiate between current and ex-IDUs, or a database that groups together individuals aged 20-29 and 30-39). Section 2.3.3 formalizes how these biases and mixture of information can be modeled.

Bayesian evidence synthesis combines prior knowledge and data from multiple sources systematically, linking different data sources through equations that describe how data sources, quantities to be estimated, and model parameters relate to each other. Two additional advantages include the ability to explicitly model the biases of each data source, and easily quantified uncertainty around the estimated quantities, due to the Bayesian approach.

We use the Bayesian evidence synthesis model developed by Sweeting et al. [Sweeting 2008; Sweeting 2009; De Angelis 2009; Harris 2012]. This model falls under the broader category of static, non-mechanistic models of infectious disease [Birrell 2017]: static because the estimates produced are for a certain point in time, and non-mechanistic in contrast to SIR models [Anderson 1991] that explicitly model the transmission mechanism. This model was first introduced for healthcare data in an evidence synthesis of HIV data [Ades 2002] and is now the UK government's key method of producing HIV prevalence estimates[2] [Hickman 2013]. We refer the reader to Ades et al. for a review of other healthcare applications of this model [Ades 2006]. The appendix provides selected mathematical details of the model from Sweeting et al. [Sweeting 2008]. The following is a summary.

## 2.3 Model

We motivated the use of Bayesian evidence synthesis to estimate disease prevalence. However, two challenges remain. First, many surveys ask only whether a respondent has ever injected drugs in their lifetime, resulting in a composite group: ever-IDUs, consisting of unknown proportions of current and ex-IDUs and lacking information on whether this ever-IDU mixture is representative of that in the general population. This makes estimating current and ex-IDU proportions $\rho_{cur,a}$ and $\rho_{ex,a}$ more complicated.

---
[2] https://www.gov.uk/government/statistics/hiv-in-the-united-kingdom

Second, current and ex-IDU data tend to be scarce compared to non-IDU data, due to injecting drug use being a sensitive risk factor typically collected through self-reporting. This makes the estimation of HCV prevalence $\pi_{cur,a}$ and $\pi_{ex,a}$ challenging.

### 2.3.1 Logistic Regressions for Data-Rich Quantities

To resolve these challenges, Sweeting et al. proposed using logistic regressions to directly model quantities for which data sources are typically more easily obtained [Sweeting 2008], namely ever-IDU proportion $\rho_{ever,a}$, ever-IDU HCV prevalence $\pi_{ever,d,tss,a}$, and non-IDU HCV prevalence $\pi_{non,a}$:

$$logit(\rho_{ever,a}) = \alpha_0 + \alpha_1 \mathbb{I}_a \quad (2)$$
$$logit(\pi_{ever,d,tss,a}) = \delta_0 + \delta_1 \mathbb{I}_a + \delta_2 \mathbb{I}_d + \delta_3 \mathbb{I}_{tss} \quad (3)$$
$$logit(\pi_{non,a}) = \gamma_0 + \gamma_1 \mathbb{I}_a \quad (4)$$

The parameters $\alpha_0, \delta_0, \gamma_0$ are scalar intercept terms. $\mathbb{I}_a$ is a vector of indicator variables for $a$ = 20-29, 30-39, or 40-49 years and $\alpha_1, \delta_1$, and $\gamma_1$ are parameter vectors of log-odd ratios of $a$ relative to the baseline of 50-59 years.

Injecting duration (D) captures the number of years an individual has injected drugs, and time since starting injecting (TSS) represents how long since an individual first started injecting. $\mathbb{I}_d$ is a vector of indicator variables for D=$d$, with time categories <1, 1-4, 5-9, 10-14, 15-19, or 20-29 years; $\mathbb{I}_{tss}$ is for TSS=$tss$ with the same categories. The parameters $\delta_2$ and $\delta_3$ are vectors of log-odd ratios of $d$ and $tss$ respectively relative to the baseline of 30-45 years. We use non-informative $Normal(0, 100)$ priors for all regression parameters $(\alpha_0, \alpha_1, \delta_0, \delta_1, \delta_2, \delta_3, \gamma_0, \gamma_1)$.

The three quantities D, TSS, and age at first use (AAFU) capture an individual's drug use history and are needed to link the data-rich quantities to other quantities to be estimated, as detailed in Section 2.3.2 below. We use ten age groups for age at first use: 8-9, 10-14, 15-19, 20-24, 25-29, 30-34, 35-39, 40-44, 45-50, 51-55 years.

### 2.3.2 Linking Data-Scarce Quantities to Data-Rich Quantities

Sweeting et al. linked quantities for which data were scarce to the data-rich quantities above [Sweeting 2008]. These data-scarce quantities are: current IDU proportion $\rho_{cur,a}$ and HCV prevalence $\pi_{cur,a}$, ex-IDU proportion $\rho_{ex,a}$ and HCV prevalence $\pi_{ex,a}$. We model the ever-IDU proportion $\rho_{ever,a}$ as a mixture of the current IDU proportion $\rho_{cur,a}$ and ex-IDU proportion $\rho_{ex,a}$, resolving the first challenge of many data sources not differentiating between current and previous injecting drug use:

$$\rho_{ex,a} = \rho_{ever,a} \kappa_{ex,a} \quad (5)$$

$$\rho_{cur,a} = \rho_{ever,a}(1 - \kappa_{ex,a}) \quad (6)$$

The parameter $\kappa_{ex,a}$, detailed in the Appendix, is the probability of being an ex-IDU conditional on being an ever-IDU, reflecting the process in which a current IDU transitions to an ex-IDU. Given the proportion of ever-IDUs $\rho_{ever,a}$, we can then calculate the proportion of non-IDUs $\rho_{non,a}$ as:

$$\rho_{non,a} = 1 - \rho_{ever,a} \qquad (7)$$

Similarly, Sweeting et al. determined current IDU HCV prevalence $\pi_{cur,a}$ and ex-IDU HCV prevalence $\pi_{ex,a}$ from ever-IDU HCV prevalence $\pi_{ever,d,tss,a}$ as functions of the distributions of drug use history quantities, mitigating the second challenge of scarcity of ex-IDU data by propagating information for current-IDUs to ever-IDUs and then ex-IDUs:

$$\pi_{cur,a} = \sum_{t=0}^{T} \pi_{ever,d=t,tss=t,a} f_{tss|cur}(t) \qquad (8)$$

$$\pi_{ex,a} = \sum_{t=0}^{T} \sum_{l=0}^{t-1} \pi_{ever,d=l,tss=t,a} \frac{f_{d|ever}}{F_{d|ever}} f_{tss|ex}(t) \qquad (9)$$

### 2.3.3 How Data Sources Inform Quantities and Parameters

Figure 1 illustrates how the various model parameters and quantities to be estimated connect to each other, and how they are informed by the data sources. Information provided by data source $i$ for distributions of drug use history can be expressed as a vector ($z_{i,1},…,z_{i,T}$), one element for each time category. All information provided for these distributions are assumed to be unbiased, and follow a multinomial distribution across categories with Dirichlet priors. These quantities appear as yellow circles in Figure 1.

Information provided by data source $i$ for IDU proportions and HCV prevalence can be expressed in terms of a numerator $y_{i,j}$ and denominator $n_{i,j}$, where $j = 1,…,m_i$ and $m_i$ is the total number of observations in data source $i$. Each observation $j$ also comes with information on IDU risk group $g$ and age group $a$ at a minimum; we also need injecting duration $d$ and time since starting injecting $tss$ information if the observation informs ever-IDU HCV prevalence. We assume $y_{i,j}$ is a realization from a $Binomial(n_{i,j}, \rho_{i,j})$ distribution for IDU proportions or a $Binomial(n_{i,j}, \pi_{i,j})$ distribution for HCV prevalence.

We formalize the scenarios described in Section 2.2 where the information provided by a data source, $\rho_{i,j}$ and $\pi_{i,j}$, can be unbiased, biased, or mixed (across IDU risk groups or age groups). Consider, for example, the quantity $\rho_{g,a}$. A data source $i$ providing unbiased information for this quantity can be linked to it using:

$$\rho_{i,j} = \rho_{g,a} \qquad (10)$$

When a data source provides biased information for a quantity, this bias $\beta$ is estimable if the bias term can be assumed to be additive on the logit scale[3], and unbiased information is available from another data source for the quantity:

---

[3] This implies bias $\beta$ affects only the absolute level of the quantity being estimated and not the odds ratios between age groups, injecting duration, or time since starting, the non-intercept regression parameters in the logistic regressions in Section 2.3.1.

$$logit(\rho_{i,j}) = logit(\rho_{g,a}) + \beta \qquad (11)$$

Figure 1 makes this clear: at least one quantity in each yellow shaded area must be informed by unbiased data.

There are multiple ways to model the bias term $\beta$. The most flexible bias structure for $\beta$ allows it to vary by data source, IDU risk group, and age group: $\beta = \beta_{g,a,i}$. The least flexible structure leaves it as a constant, $\beta$. Intermediate options include $\beta = \beta_{g,i}$, where bias is allowed to vary by data source and IDU risk group but is constant across age groups, and several others we consider in Section 5.1. We use non-informative $Normal(0, 100)$ priors for all bias terms.

We model data sources that mix information across IDU risk groups or age groups as mixtures of the quantities they inform:

$$\rho_{i,j} = \sum_{g \in g'} \sum_{a \in a'} \omega_{g,a} \rho_{g,a} \qquad (12)$$

where $\omega_{g,a}$ is the number of individuals of IDU $g$ and age $a$ (as a proportion) of the size of the subpopulation covered by that observation. As an example, the National Survey on Drug Use and Health (NSDUH), in which the ex-IDU proportion was combined across two age groups, can be modeled as:

$$\rho_{ex,30-49,NSDUH} = \omega_{ex,30-39}\rho_{ex,30-39} + \omega_{ex,40-49}\rho_{ex,40-49} \qquad (13)$$

with $\omega_{ex,30-39} = \frac{N_{30-39}}{N_{30-39}+N_{40-49}}$ and $\omega_{ex,40-49} = \frac{N_{40-49}}{N_{30-39}+N_{40-49}}$.

### 2.3.4 Other Quantities Estimated

In addition to overall HCV prevalence $\pi$, HCV prevalence among type $g$ IDUs aged $a$ $\pi_{g,a}$, and the proportion of type $g$ IDUs among individuals aged $a$ $\rho_{g,a}$, we also estimate the following quantities: HCV prevalence among individuals aged $a$ $\pi_a$, HCV prevalence among type $g$ IDUs $\pi_g$, and the proportion of type $g$ IDUs across all ages $\rho_g$:

$$\pi_a = \rho_{non,a}\pi_{non,a} + \rho_{cur,a}\pi_{cur,a} + \rho_{ex,a}\pi_{ex,a} \qquad (14)$$

$$\rho_g = \frac{\sum_a (N_a \rho_{g,a})}{\sum_a N_a} \qquad (15)$$

$$\pi_g = \frac{\sum_a (N_a \rho_{g,a} \pi_{g,a})}{\sum_a (N_a \rho_{g,a})} \qquad (16)$$

### 2.3.5 Role of Drug Use History Information

We need the distributions of drug use history variables D, TSS, and AAFU to relate data-scarce quantities to data-rich quantities, as detailed in Section 2.3.2. Estimating these distributions for ever-IDUs from a sample of ever-IDUs may be inaccurate, as the proportion of current to ex-IDUs in such a sample is unknown and the sample may not be representative of ever-IDUs in the population. However, estimating these ever-IDU distributions from samples of current or ex-IDUs without further adjustments does not resolve the issue either – since long-term users are more likely to still be injecting, a sample of ex-IDUs will over-represent short-term users, resulting in low estimates of D and TSS [Kaplan 1997]; likewise, a sample of current IDUs will over-represent long-term users, resulting in high estimates of D and TSS.

To account for these biases, Sweeting et al. used an adjustment method proposed by Kaplan [Kaplan 1997] to yield unbiased estimates of $f_{D|ever}$ and $f_{TSS|ever}$ for ever-IDUs starting from ex-IDU data. The Appendix details this adjustment method. We use both current and ex-IDU data for TSS and ex-IDU data for D to inform the estimation of $f_{D|ever}$ and $f_{TSS|ever}$.

### 2.3.6 New Equations to Adjust $f_{aafu|cur}$ and $f_{aafu|ex}$ for $f_{aafu|ever}$

In Sweeting et al., the AAFU distribution for ever-IDUs was informed by data from a program that provides treatment services to ever-IDUs. However, for the same reasons described above – unknown proportion of current to ex-IDUs in a sample of ever-IDUs, and lack of information on how representative the sample of ever-IDUs is of the population of ever-IDUs – estimates of the AAFU distribution for ever-IDUs from a sample of ever-IDUs may be inaccurate. To avoid this potential inaccuracy, we derive two new equations to adjust for the same bias in AAFU data for current and ex-IDUs, allowing these data to be incorporated into the model. Detailed proofs are in the Appendix.

$$f_{AAFU|cur,a}(a-t) = \frac{f_{AAFU|ever,a}(a-t)[1-F_{D|ever,TSS=t,a}(t)]}{1-\kappa_{ex,a}} \quad (17)$$

$$f_{AAFU|ex,a}(a-t) = \frac{f_{AAFU|ever,a}(a-t)F_{D|ever,TSS=t,a}(t)}{\kappa_{ex,a}} \quad (18)$$

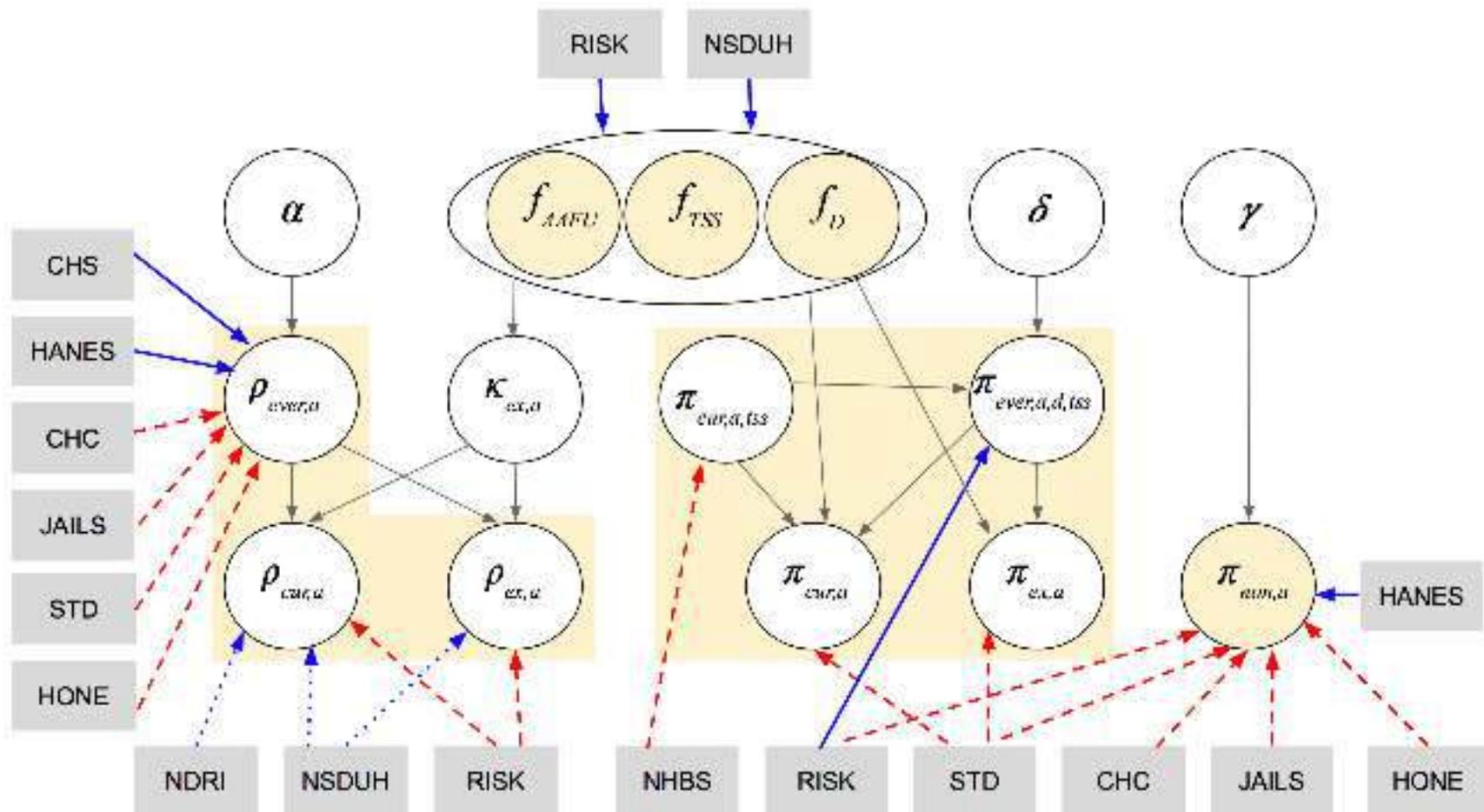

**Figure 1** The relationship between model parameters, quantities to be estimated, and data sources. Data sources are in boxes; parameters and quantities are in circles. A blue line represents unbiased information provided by a data source for a quantity; a red line represents biased information; a dotted line represents information mixed across age groups. For the model to be estimable, unbiased information is required for all quantities in yellow circles and at least one quantity in each yellow shaded box.

CHS: Community Health Survey; HANES: Health and Nutrition Examination Survey; CHC: Check hepatitis C Program; JAILS: Correctional Health Services hepatitis C Serosurvey; STD: Sexually Transmitted Disease Control Clinics; HONE: Hepatitis Outreach Network; NDRI: National Development and Research Institutes for

## 3 Data

A data source provides either unbiased, biased, both unbiased and biased, or mixed (across IDU risk groups or age groups) information for IDU proportions and HCV prevalence. Some data sources also provide information on the distributions of drug use history variables. We use ten primary data sources, where eight are for NYC, one is for the New York Metropolitan Statistical Area (NYMSA), and one is national. We also use two additional data sources to provide auxiliary information needed by the model, such as the size of the NYC population stratified by age, and multipliers to adjust national and NYMSA data for NYC.

We adjust data from sources measured at the national or NYMSA levels to reflect the IDU risk group and age group compositions in NYC. To obtain multipliers for IDU risk groups, we calculate the proportion of NYC residents among both NYMSA patients and patients nationwide from the National Survey of Substance Abuse Treatment Services, a census of all substance abuse treatment facilities in the US. For age group multipliers, we use the proportion of the NYC population in each age group from the US Census. We then multiply national and NYMSA estimates by the IDU risk group and age group multipliers to obtain NYC-level estimates.

Table 1 summarizes the quantities informed by the data sources. The majority of data were collected between 2005 and 2012, hence we assume the resulting prevalence estimates are applicable for the year 2012, since HCV prevalence is likely to be fairly stable in the short term.

**Table 1**      Information provided by data sources

| Data Source | Year | Reference | Information Type | Quantities Informed |
|---|---|---|---|---|
| NYC Check Hepatitis C Program (CHC) | 2012 | [CHC 2012] | Biased | $\rho_{ever,a}$, $\pi_{non,a}$ |
| NYC Community Health Survey (CHS) | 2003 | [CHS 2003] | Unbiased | $\rho_{ever,a}$ |
| NYC Health and Nutrition Examination Survey (HANES) | 2004 | [HANES 2004] | Unbiased | $\rho_{ever,a}$, $\pi_{non,a}$ |
| NYC Hepatitis Outreach Network (HONE) | 2009-2011 | [HONE 2009] | Biased | $\rho_{ever,a}$, $\pi_{non,a}$ |
| NYC Correctional Health Services Hepatitis C Serosurvey (JAILS) | 2006 | [JAILS 2006], [Parvez 2016] | Biased | $\rho_{ever,30-39}$, $\pi_{non,30-39}$ |
| National Development and Research Institutes (NDRI) for NYMSA | 2007 | [Tempalski 2013] | Mixture across age groups | $\rho_{cur,20-59}$ |
| National HIV Behavioral Surveillance (NHBS) for NYC | 2009 | [Lansky 2007], [NYCDOHMH 2010] | Biased | $\pi_{cur,a,tss}$ |
| National Survey on Drug Use and Health (NSDUH) | 2011 | [NSDUH 2011] | Mixture across age groups | $\rho_{cur,30-49}$, $\rho_{ex,30-49}$, $f_D$, $f_{TSS}$, $f_{AAFU}$ |

| Risk Factors for HIV/AIDS among NYC IDUs (RISK) | 2005-2012 | [Jordan 2015] | Both | $\rho_{cur,a}$, $\rho_{ex,a}$, $\pi_{ever,a,d,tss}$, $\pi_{non,a}$, $f_D$, $f_{TSS}$, $f_{AAFU}$ |
|---|---|---|---|---|
| NYC Sexually Transmitted Disease Control Clinics (STD) | 2005-2010 | [STD 2005] | Biased | $\rho_{ever,a}$, $\pi_{cur,a}$, $\pi_{ex,a}$, $\pi_{non,a}$ |
| US Census | 2010 | [Census 2010] | Auxiliary | NYC population by age group |
| National Survey of Substance Abuse Treatment Services | 2007, 2011 | [N-SSATS 2007, N-SSATS 2011] | Auxiliary | Proportion of patients nationwide/NY MSA from NYC |

In general, we consider unbiased information to be that provided by data sources based on random samples. Data from specialized subpopulations, such as the incarcerated, immigrant communities, or homeless individuals, are unlikely to be representative of NYC residents as a whole, hence we consider them biased; the same applies to data based on individuals who seek out specific services such as low-cost healthcare for chronic diseases, or data collected using respondent-driven methods or similar non-probability sampling methods. The exception is the RISK data. This data was collected from subjects in an inpatient drug treatment program at Mount Sinai Beth Israel Hospital in NYC. We expect the information it provides on IDU proportions to be biased, since as a drug treatment program, it may have a higher current to ex-IDU proportion compared to the NYC population. On the other hand, we consider the HCV prevalence and drug use history of its IDUs to be reflective of IDUs in general.

*Unbiased data on proportion of injecting drug users.* Four data sources provide unbiased information on IDU proportions. Two of them, CHS and HANES, are household surveys designed to be representative of the NYC population. They survey non-institutionalized adults (20 years and older) in NYC, where lifetime injecting drug use is self-reported. NDRI, a nonprofit organization studying substance abuse, completed an evidence synthesis model that generates national estimates of current IDU prevalence and allocates them across metropolitan areas using multiplier methods [Tempalski 2013]. We adjust their NYMSA estimate to the NYC level as above. The NSDUH survey provides self-reported data on current and lifetime injecting drug use from a random sample of individuals nationwide. We adjust this national estimate to the NYC level as described above. Information provided by the last two data sources is mixed across age groups (20-59 years old for NDRI, 30-49 years old for NSDUH).

*Biased data on proportion of injecting drug users.* Five data sources provide biased information on IDU proportions. Four inform ever-IDU proportion; we consider them biased because they are based on specialized subpopulations that likely have a higher proportion of IDUs compared to the NYC population. They include: the CHC program that diagnoses and treats individuals with chronic HCV, covering low-cost community health centers and syringe exchange facilities in neighborhoods with high rates of newly reported HCV; HONE, a similar community health program ran by NYC's

Mount Sinai Medical Center for immigrants from foreign countries where Hepatitis B and C are endemic; data from patients who visit the STD clinics for low-cost STD testing and treatment; and JAILS data from individuals newly incarcerated in seven NYC jails. The CHC data source includes homeless individuals. The fifth data source, RISK, informs current and ex-IDU proportions, which we consider biased as above. Injecting drug use is self-reported in all data sources.

*Data on hepatitis C prevalence.* We have one source of unbiased information for HCV prevalence (HANES), one source that provides both unbiased and biased information depending on IDU risk group (RISK), and five sources that provide biased information. All perform serological testing for HCV. We expect the non-IDUs surveyed by HANES, a household survey of NYC adults, to be representative of non-IDUs in the general population. As a drug treatment program, we expect the information provided by RISK for HCV prevalence in IDUs to be unbiased, and non-IDUs to be biased – non-IDUs in this data are current or ex drug users who are using or used non-injecting forms of drugs.

Four more sources also provide HCV prevalence information for non-IDUs: CHC, HONE, STD, and JAILS, the same data sources whose IDU size information we considered biased above. Non-IDUs in these specialized subpopulations likely experience other HCV risk factors not experienced by their counterparts in the general NYC population, hence we consider their information biased. For the same reason, we also treat the STD clinics' current and ex-IDUs HCV prevalence as biased. Finally, the NHBS data, based on a respondent-driven sample of IDUs, men who have sex with men, and heterosexuals at high risk for HIV, has information on HCV prevalence among current IDUs stratified by age group and time since starting. We consider this information biased for several reasons, including NHBS' respondent-driven data collection and likely HIV/HCV co-infection mechanics.

*Data on drug use history.* We obtain information on distributions of injecting duration, time since starting injecting, and age at first use for current, ex, and ever-IDUs from NSDUH and RISK.

# 4 Implementation and Evaluation

We use WinBUGS [Lunn 2001] to perform Bayesian inference using Markov chain Monte Carlo methods and obtain posterior distributions of the model parameters and quantities to be estimated. We use posterior means as point estimates and the 2.5 and 97.5 percentiles to form 95% credible intervals as a measure of uncertainty. Unless otherwise noted, results below are based on 46,000 iterations per chain of 2 parallel chains. We discard the first 4,000 iterations of each chain as burn-in and assess model convergence using trace plots, kernel density plots, and the Gelman-Rubin diagnostic [Gelman 1992].

To assess model goodness-of-fit, we use standardized deviance. For each data source $i$ with $j = 1, \ldots, m_i$ observations where the model predicts IDU proportion $\hat{\rho}_{i,j}$ or HCV prevalence $\hat{\pi}_{i,j}$, we calculate the standardized deviance using the likelihood for independent binomial random variables [McCullagh 1989]:

$$\Delta_i = 2 \sum_{j=1}^{m_i} (y_{i,j} \log \frac{y_{i,j}}{n_{i,j}\hat{\rho}_{i,j}} + (n_{i,j} - y_{i,j}) \log \frac{n_{i,j} - y_{i,j}}{n_{i,j} - n_{i,j}\hat{\rho}_{i,j}}) \quad (19)$$

Lower values indicate better goodness-of-fit. For each data source $i$ where the model predicts the distribution of a drug use history quantity ($\hat{z}_{i,1},..., \hat{z}_{i,T}$), we calculate standardized deviance using the likelihood for multinomial random variables:

$$\Delta_i = 2 \sum_{t=1}^{T} (z_{i,t} \, log \, \frac{z_{i,t}}{\hat{z}_{i,t}}) \qquad (20)$$

Since data sources are assumed to be independent of each other, a model's standardized deviance can be computed by summing the deviance contributed by each of its data sources:

$$\Delta_{model} = \sum_{i \, \in \, model} \Delta_i \qquad (21)$$

We used means of the posterior distributions of the deviance terms as their point estimates. As Sweeting et al. note, a model's posterior mean deviance is not, on its own, a reliable indicator of *absolute* goodness-of-fit when $n_{i,j}$ is small [McCullagh 1989], as is the case in many of the data sources. However, *relative* goodness-of-fit can be determined from how a data source's deviance, or a group of deviances, changes between models. Two important considerations when building the model are: first, which data source is assumed to provide unbiased information for which parameter; second, the assumed bias structure (c.f. Section 2.3.3) for the data source assumed to provide biased information. Choices made for these two considerations result in a particular bias formulation. We search for the best-fitting bias formulation by examining the change in deviance when bias formulations vary between models. We also use this criterion to investigate the model's goodness-of-fit, using the change in deviance of a data source between models to check the consistency between data sources.

## 5  Results

We first present estimates for IDU proportions and HCV prevalence from our chosen model. We compare these estimates to existing estimates in Section 6.1. To address the modeling considerations described above, in Section 5.1 we describe the various bias formulations we considered before choosing this model, and in Section 5.2 we investigate the chosen model's goodness-of-fit and consistency. Our chosen model, Model B5, uses data source and IDU risk group specific bias and has the lowest overall deviance.Figure 1 and Table 1 detail which data source it considers biased or unbiased.

We present estimates from the chosen model (Model B5) for IDU proportions and HCV prevalence by IDU risk groups and age groups, as well as the associated uncertainty. Table 2 provides estimates of injecting drug use in the NYC population by age group. Among NYC adults aged 20-29, an estimated 0.76% (10,400 individuals) are current IDUs and 1.42% (19,500 individuals) are ex-IDUs, for a total ever-IDU population aged 20-29 of 2.18%, or 29,900 individuals. The last row in Table 2 summarizes injecting drug use across all age groups. An estimated 0.58% of adults aged 20-59 are currently injecting drugs, amounting to 27,600 individuals. Of these, 2.53% (120,700 individuals) are ex-IDUs, for a total of 3.11% (148,200 individuals) of the NYC 20-59 population that has ever injected drugs in their lifetime.

Injecting drug use varies by age. Older subpopulations have greater proportions of individuals who have ever injected drugs (3.95% of adults aged 50-59 compared to the 2.18% of NYC adults aged 20-29 mentioned above; Table 3, last column). However, the older an ever-IDU, the more likely it is that he or she has stopped injecting. This can be seen from the proportion of ever-IDUs who are ex-IDUs in each age group: 65%, 75%, 85%, and 96% for ages 20-29, 30-39, 40-49, and 50-59 respectively. Thus, younger IDUs are more likely to still be injecting, which we see in the reverse patterns of injecting drug use by age among ex-IDUs in Figure 2 and current IDUs in Figure 3.

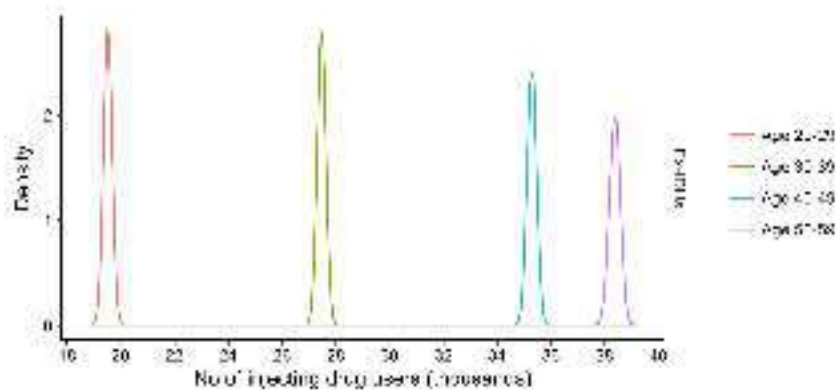

**Figure 2** Posterior distributions of number of ex-IDUs in each age group

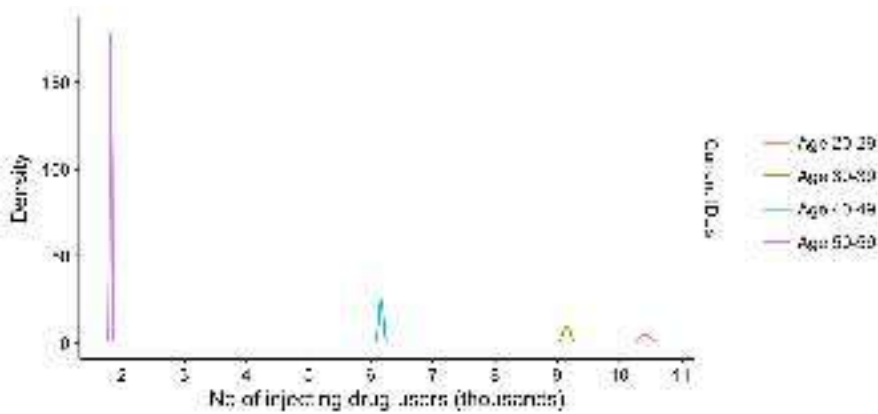

**Figure 3** Posterior distributions of number of current IDUs in each age group

Table 3 shows estimated HCV prevalence in the NYC population by IDU risk group and age group. We estimate that 0.27% of NYC adults aged 20-29 (3,700 individuals) are HCV-positive current IDUs, and 0.42% (5,700 individuals) are HCV-positive ex-IDUs.

The last row in Table 3 summarizes HCV prevalence in each IDU risk group. An estimated 0.29% (13,800 individuals) of the NYC population aged 20-59 are currently injecting drugs and are HCV positive. These 13,800 individuals make up 50.2% (95% CI 47.0% – 53.4%) of the 27,600 current IDUs in NYC (Table 2, last row). This high percentage is driven by two factors: the high numerator dominated by 5,100 current IDUs aged 30-39 who are HCV positive, and the low denominator of 27,600 current IDUs in NYC (relative to NYC population numbers). Similarly for ex-IDUs, an estimated 1.48% of the NYC population aged 20-59, or 70,500 individuals, injected drugs in the past and are HCV positive. These individuals are 58.5% (95% CI 52.3% – 64.5%) of the

**Table 2** Number of current and ex-IDUs by age group

| Age Group | NYC Population Size (1000s) | Current IDUs | | Ex-IDUs | | Ever-IDUs* | |
|---|---|---|---|---|---|---|---|
| | | N (1000s) | % | N (1000s) | % | N (1000s) | % |
| 20-29 | 1,372.775 | 10.4 (10.2 – 10.6) | 0.76 (0.74 – 0.78) | 19.5 (19.2 – 19.9) | 1.42 (1.39 – 1.45) | 29.9 (29.4 – 30.5) | 2.18 (2.14 – 2.22) |
| 30-39 | 1,249.662 | 9.2 (9.0 – 9.3) | 0.73 (0.72 – 0.75) | 27.5 (27.1 – 27.8) | 2.20 (2.17 – 2.23) | 36.6 (36.2 – 37.1) | 2.93 (2.89 – 2.97) |
| 40-49 | 1,132.972 | 6.2 (6.0 – 6.3) | 0.54 (0.53 – 0.55) | 35.3 (34.9 – 35.7) | 3.12 (3.08 – 3.15) | 41.5 (41.0 – 41.9) | 3.66 (3.62 – 3.70) |
| 50-59 | 1,017.219 | 1.8 (1.7 – 1.9) | 0.18 (0.17 – 0.19) | 38.4 (37.9 – 38.8) | 3.77 (3.73 – 3.82) | 40.2 (39.7 – 40.7) | 3.95 (3.90 – 4.00) |
| Total* (20-59) | 4,772.628 | 27.6 (27.3 – 27.8) | 0.58 (0.57 – 0.59) | 120.7 (120.0 – 121.2) | 2.53 (2.51 – 2.54) | 148.2 (146.4 – 150.0) | 3.11 (3.06 – 3.15) |

* numbers may not total due to rounding. % calculated by multiplying $\rho_{g,a}$ by 100. 95% credible interval in parentheses. NYC subpopulation sizes stratified by age group obtained from the US Census [Census 2010]. Ever-IDUs include current and ex-IDUs. Non-IDU estimated percentages can be calculated as 1 – Ever-IDU estimated percentages.

120,700 estimated ex-IDUs in NYC (Table 2, last row). Finally, the model estimates 1.01% (48,200 individuals) of the NYC population aged 20-59 to be non-IDUs who are HCV positive. They are 1.04% (95% CI 1.03% – 1.06%) of the 4,624,000 non-IDUs in NYC.

The most affected group is ex-IDUs aged 50-59, with 29,100 individuals in this group (2.86% of the NYC 50-59 population) being HCV positive. Across all age groups, HCV prevalence is highest for ex-IDUs compared to current and non-IDUs. The HCV estimates for ex-IDUs also have the highest uncertainty, as can be seen from their larger credible intervals in Table 3 and less sharp posterior distributions in Figure 4 compared to other IDU risk groups.

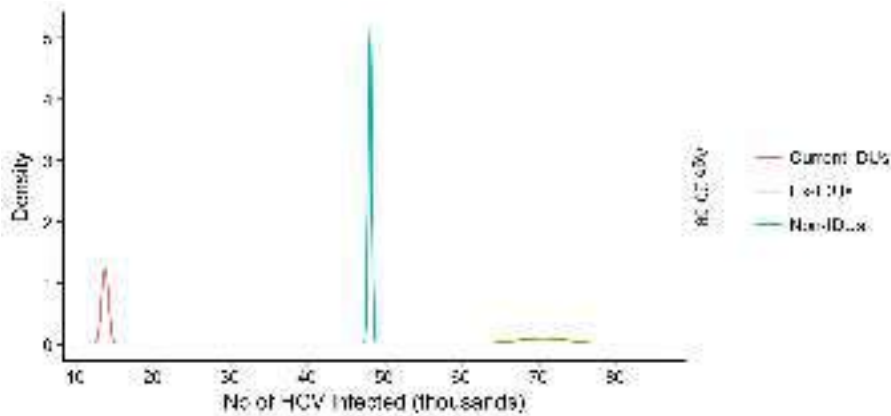

**Figure 4**     Posterior distributions of number of HCV infected for each IDU risk group

The high rates of infection among older non-IDUs aged 40-59 may be attributed to the dynamics of HCV transmission among this generation of individuals born between 1953 and 1972, or "baby boomers". While these dynamics are not completely understood, several studies have pointed to unsafe medical procedures of that time and contaminated blood before widespread blood screening procedures as drivers of HCV transmission in baby boomers [Spaulding 2016; Ward 2013].

Just as injecting drug use varies by age, HCV prevalence also varies by age. Ignoring IDU risk group, a greater percentage of older subpopulations are HCV positive. From Table 4, adults aged 50-59 (who were born in 1953-1962) have the highest HCV positive rate of 5.27%, amounting to 536,000 individuals. However, this increase in HCV prevalence with age does not hold across all injecting drug users. Notably, this is not the case for current injecting drug users – among younger NYC adults (20-39 years old), current injecting drug users have higher HCV prevalence than their ex- and non-IDU counterparts (Figures 4 and 5), agreeing with recent studies reporting increasing numbers of new HCV infections among young injecting drug users in New York state [CDC 2008] and other regions in the US [Suryaprasad 2014; CDC 2011; CDC 2015].

After weighting HCV prevalence estimates by subpopulation sizes, the model estimates overall HCV prevalence among the NYC 20-59 population to be 2.78% (95% CI 2.61% to 2.94%), or 132,500 individuals (95% CI 124,900 to 140,000 individuals) (Table 4, last row).

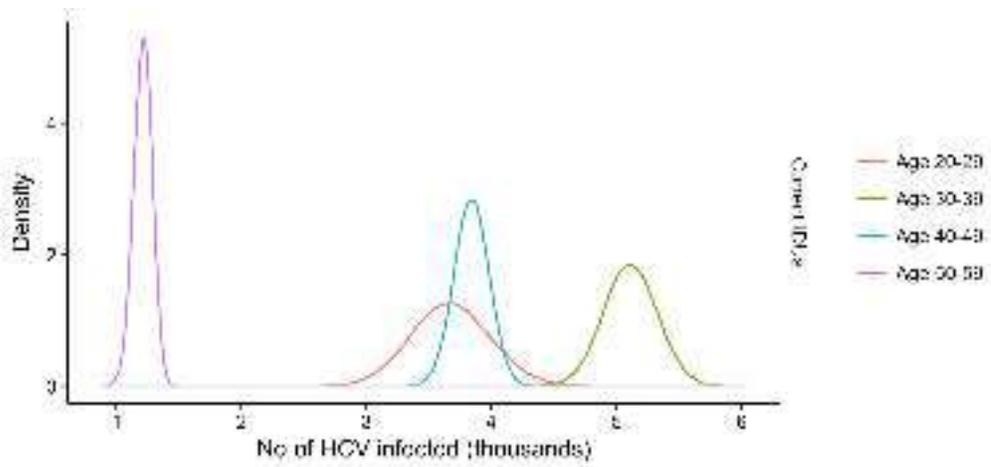

**Figure 5**   Posterior distributions of number of HCV infected current IDUs

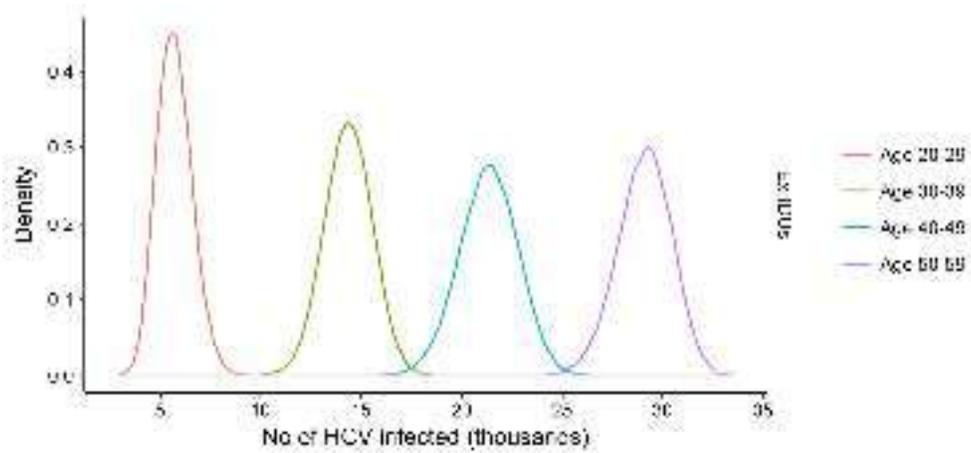

**Figure 6**   Posterior distributions of number of HCV infected ex-IDUs

Table 3  Number of HCV infected by IDU risk group and age group

| Age Group | NYC Population Size (1000s) | Current IDUs | | Ex-IDUs | | Non-IDUs | |
|---|---|---|---|---|---|---|---|
| | | N (1000s) | % | N (1000s) | % | N (1000s) | % |
| 20-29 | 1,372.775 | 3.7 (3.0 – 4.4) | 0.27 (0.21 – 0.32) | 5.7 (4.0 – 7.7) | 0.42 (0.29 – 0.56) | 0 (0.0 – 0.1) | 0 (0 – 0.01) |
| 30-39 | 1,249.662 | 5.1 (4.6 – 5.6) | 0.41 (0.37 – 0.45) | 14.4 (11.8 – 16.9) | 1.15 (0.95 – 1.35) | 5.0 (4.8 – 5.2) | 0.40 (0.39 – 0.42) |
| 40-49 | 1,132.972 | 3.8 (3.5 – 4.2) | 0.34 (0.31 – 0.37) | 21.3 (18.2 – 24.4) | 1.88 (1.61 – 2.15) | 19.8 (19.5 – 20.1) | 1.75 (1.72 – 1.78) |
| 50-59 | 1,017.219 | 1.2 (1.0 – 1.4) | 0.12 (0.10 – 0.14) | 29.1 (26.0 – 31.9) | 2.86 (2.56 – 3.14) | 23.3 (22.9 – 23.7) | 2.29 (2.26 – 2.33) |
| Total* (20-59) | 4,772.628 | 13.8 (12.8 – 14.8) | 0.29 (0.27 – 0.32) | 70.5 (62.8 – 78.1) | 1.48 (1.31 – 1.64) | 48.2 (47.7 – 48.7) | 1.01 (1.00 – 1.02) |

\* numbers may not total due to rounding. % calculated by multiplying $\rho_{g,a}$ by $\pi_{g,a}$ and 100. 95% credible interval in parentheses. NYC subpopulation sizes stratified by age group obtained from the US Census [Census 2010].

Table 4  Number of HCV infected by age group

| Age Group | NYC Population Size (1000s) | All IDU Risk Groups | |
|---|---|---|---|
| | | N (1000s) | % |
| 20-29 | 1,372.775 | 9.4 (7.4 – 11.7) | 0.69 (0.54 – 0.85) |
| 30-39 | 1,249.662 | 24.5 (21.9 – 27.1) | 1.96 (1.75 – 2.17) |
| 40-49 | 1,132.972 | 45.0 (42.0 – 47.9) | 3.97 (3.71 – 4.23) |
| 50-59 | 1,017.219 | 53.6 (50.8 – 56.2) | 5.27 (4.99 – 5.53) |
| Total* (20-59) | 4,772.628 | 132.5 (124.9 – 140.0) | 2.78 (2.61 – 2.94) |

## 5.1 Bias Modeling

In this section, we describe the various bias formulations we considered before choosing Model B5, whose results we presented in the previous section. Table 5 presents seven models that make different assumptions about which types of information are considered biased and how biases vary in data sources.

Model B1 assumes no information provided is biased. Model B5 has the most flexible bias structure, with bias allowed to vary by data source and IDU risk group. Models B2-4 are intermediaries of Model B1 and B5, relaxing Model B1's no-bias assumption for one IDU risk group at a time. For example, the following information for current IDUs is assumed to be biased in Model B5, but unbiased in Model B2: RISK data's current IDU proportion, NHBS data's current IDU HCV prevalence with time since starting information, and STD data's current IDU HCV prevalence. Similar scenarios apply for Model B3 and B4, but for ex- and non-IDUs. We also explore two additional bias structures less flexible than that of Model B5. In Model B6, bias terms are allowed to vary across IDU risk group but not data source; in Model B7, bias terms are allowed to vary across data source but not IDU risk group.

Of the ten primary data sources, six (CHC, JAILS, STD, HONE, NHBS, RISK – for quantities $\rho_{cur,a}$, $\rho_{ex,a}$ and $\pi_{non,a}$) are biased in Model B5 but are set to unbiased or have different bias structures in other models. Model deviance $\Delta_{model}$ can then be decomposed into two groups, $\Delta_{b \to ub}$ and $\Delta_{ub}$, where $\Delta_{b \to ub}$ sums deviance contributions of the six data sources above, and $\Delta_{ub}$ sums the deviance contributions of the remaining four primary data sources (CHS, HANES, NDRI, NSDUH) set to unbiased in all models.

**Table 5** Goodness-of-fit of models with different bias formulations

| Model | Description | $\bar{\Delta}_{model}$ | $\bar{\Delta}_{ub}$ | $\bar{\Delta}_{b \to ub}$ |
|---|---|---|---|---|
| B1 | No bias, $\beta_{g,i} = 0$ | 142,485 | 124,851 | 17,634 |
| B2 | Current IDU no bias, $\beta_{cur,i} = 0$ | 131,020 | 124,822 | 6,198 |
| B3 | Ex-IDU no bias, $\beta_{ex,i} = 0$ | 130,424 | 124,827 | 5,597 |
| B4 | Non-IDU no bias, $\beta_{non,i} = 0$ | 139,865 | 124,828 | 15,037 |
| B5 | Data source and IDU risk group specific bias, $\beta_{g,i}$ | 129,959 | 124,814 | 5,145 |
| B6 | IDU risk group specific bias, $\beta_g$ | 131,958 | 124,821 | 7,137 |
| B7 | Data source specific bias, $\beta_i$ | 133,284 | 124,824 | 8,460 |

$\bar{\Delta}_{model}$ = posterior mean deviance. $\bar{\Delta}_{ub}$ = posterior mean deviance of unbiased data sources. $\bar{\Delta}_{b \to ub}$ = posterior mean deviance of data sources treated as biased in Model B5 but are set to unbiased or have different bias structures in other models. $\bar{\Delta}_{model} = \bar{\Delta}_{ub} + \bar{\Delta}_{b \to ub}$

All models besides Model B5 are based on 6,000 iterations per chain of 2 parallel chains. Model B5 uses 46,000 iterations per chain. The first 4,000 iterations of all chains were discarded as burn-in.

Model B5 has the lowest overall deviance (Table 5). The deviance contributions of data sources set as unbiased in all bias formulations, $\bar{\Delta}_{ub}$, remains at similar levels throughout. However, $\bar{\Delta}_{b \to ub}$ varies significantly across bias formulations. Assuming that no data sources are biased returns a model (Model B1) with the largest deviance, driven by a large increase in $\bar{\Delta}_{b \to ub}$. A close look at the data sources making up $\bar{\Delta}_{b \to ub}$ revealed that the JAIL data is the least well-fit when forced to be unbiased – its deviance increased 77.5 times compared to its deviance in Model B5. The Check Hep C program had an increase of 38 times. The HONE and NHBS data had less dramatic increases, with HONE increasing by 1.5 times and NHBS data by 1.4 times. After Model B1, Model B4 where non-IDUs from high-risk subpopulations such as the incarcerated, immigrant communities, or homeless individuals are considered representative of non-IDUs in the NYC population, has the largest $\bar{\Delta}_{b \to ub}$. These results suggest that differential bias for non-IDUs at a minimum is needed.

We also consider the epidemiological implications of the different bias formulations by examining the quantities estimated (Table 6). Overall HCV prevalence is highest in Model B4. This is unsurprising, as this model assumes non-IDU data from specialized, high-risk subpopulations is unbiased. Across the different bias formulations, the largest changes in estimates occur for HCV prevalence in current IDUs. IDU proportions $\rho_{cur}$ and $\rho_{ex}$ do not change much as the bias formulation changes, suggesting that data sources for IDU proportions that are considered biased in Model B5 but unbiased or modeled differently in other models are not influential. Model B7, in which bias was not allowed to vary by IDU risk group, had the lowest overall HCV prevalence. Model B7 was also the only model besides Model B1 where all information was considered unbiased, without differential bias by IDU risk group. Unlike the rest of the models, these two models estimate HCV prevalence in current IDUs to be higher than that of ex-IDUs, which may point towards imprecise separation of $\pi_{ever,a,d,tss}$ data into HCV prevalence among current and ex-IDUs. Models B5 and B6 provide similar estimates, so we refer back to their goodness-of-fit statistics. That deviance decreases when moving from Model B6 to Model B5 suggests that suggests differential bias by data source can be helpful. Hence, we select Model B5's bias formulation – differential bias by data source and IDU risk group.

**Table 6** Quantities estimated by models with different bias formulations

| Model | Description | $\rho_{cur}$ | $\rho_{ex}$ | $\pi_{cur}$ | $\pi_{ex}$ | $\pi_{non}$ | $\pi$ |
|---|---|---|---|---|---|---|---|
| B1 | No bias, $\beta_{g,i} = 0$ | 0.58 (0.57 – 0.59) | 2.55 (2.53 – 2.56) | **59.2** (56.4 – 61.9) | **54.7** (49.8 – 59.6) | 1.05 (1.04 – 1.06) | 2.76 (2.62 – 2.89) |
| B2 | Current IDU no bias, $\beta_{cur,i} = 0$ | 0.58 (0.57 – 0.59) | 2.53 (2.51 – 2.54) | 50.2 (47.1 – 53.4) | 58.5 (52.2 – 64.5) | 1.05 (1.04 – 1.07) | 2.79 (2.62 – 2.95) |
| B3 | Ex-IDU no bias, $\beta_{ex,i} = 0$ | 0.58 (0.57 – 0.59) | 2.53 (2.52 – 2.55) | 49.9 (46.8 – 53.0) | 53.5 (48.5 – 58.4) | 1.04 (1.03 – 1.06) | 2.65 (2.52 – 2.78) |
| B4 | Non-IDU no bias, $\beta_{non,i} = 0$ | 0.58 (0.57 – 0.59) | 2.54 (2.52 – 2.55) | 59.5 (56.7 – 62.3) | 60.5 (54.4 – 66.7) | 1.04 (1.03 – 1.06) | **2.89** (2.73 – 3.05) |
| B5 | Data source and IDU risk group specific bias, $\beta_{g,i}$ | 0.58 (0.57 – 0.59) | 2.53 (2.51 – 2.54) | 50.2 (47.0 – 53.4) | 58.5 (52.3 – 64.5) | 1.04 (1.03 – 1.06) | 2.78 (2.61 – 2.94) |
| B6 | IDU risk group specific bias, $\beta_g$ | 0.58 (0.57 – 0.58) | 2.53 (2.51 – 2.54) | 47.1 (44.1 – 50.1) | 57.8 (51.6 – 63.7) | 1.04 (1.03 – 1.06) | 2.74 (2.58 – 2.90) |
| B7 | Data source specific bias, $\beta_i$ | 0.58 (0.57 – 0.59) | 2.52 (2.51 – 2.54) | **45.9** (43.1 – 49.0) | **43.7** (38.7 – 48.8) | 1.04 (1.03 – 1.06) | **2.38** (2.25 – 2.52) |

All models besides Model B5 are based on 6,000 iterations per chain of 2 parallel chains. Model B5 uses 46,000 iterations per chain. The first 4,000 iterations of all chains were discarded as burn-in.

## 5.2 Goodness-Of-Fit and Model Consistency

We now study the goodness-of-fit of the chosen model (Model B5). To determine whether a data source is influential for model estimates or is in conflict with other data sources, we follow the cross-validation strategy proposed by Sweeting et al., removing one data source at a time, and re-fitting the model [Sweeting 2008]. If the deviance of a data source decreases when another data source is removed from the model, this suggests the two data sources are potentially in conflict. Each row in Table 7 reports the deviance estimates under the re-fitted model, and Table 8 reports the prevalence quantities estimated.

Currently, the IDU and HCV quantities (for current and ex-IDUs only) are connected through the drug use history variables. Hence, information on IDU proportions could affect HCV prevalence, and vice versa. When the HANES data or RISK data are removed, several IDU and HCV prevalence estimates do not converge (Table 8). Without RISK, the model does not receive any unbiased information for HCV prevalence quantities hence the current and ex-IDUs HCV prevalence quantities do not converge. Similarly, HANES is the only source of unbiased information for non-IDU HCV prevalence, and this estimate does not converge without it. We also assume that HANES informs ever-injecting drug use with no bias; without it, the ex-IDU quantities do not converge. The current-IDU proportion estimate still converges, perhaps because it is additionally informed by the NDRI data. While HANES does not inform ex-IDU HCV prevalence, that ex-IDU HCV prevalence does not converge with its removal is evidence that information is being propagated from IDU quantities to HCV quantities.

The deviance of the NHBS data decreases when the RISK data is removed (Table 7) and vice versa, yet including both is necessary to improve how ever-IDU HCV prevalence is attributed to current and ex-IDUs. When the NHBS data is removed, current IDU HCV prevalence increases by 5%, and ex-IDU prevalence increases by 1% (Table 8). Aside from RISK and NHBS, only the STD data collected on a high-risk subpopulation provides (biased) information for current and ex-IDUs HCV prevalence. However, the STD data is not influential, with HCV prevalence scarcely changing when it is removed. Of all HCV prevalence quantities, those for ex-IDUs are the most uncertain (Table 3; Figure 4). Hence, we retain the NHBS data in the analysis to avoid further increasing uncertainty of these HCV prevalence quantities, and emphasize that more HCV prevalence data for current or ex-IDUs is needed to improve the precision of these estimates.

Finally, we also check the assumed prior distributions. The model in the first row of Table 8 is solely informed by its priors without any data sources, serving as a baseline for whether the prior distributions were sufficiently non-informative as is desired in this setting where we wish for the data sources to drive the results. The HCV prevalence estimates are around 0.5, as expected.

### 4.3.1 Alternative Treatments of IDU Proportion Data Sources

Some conflict is present between three of the four data sources assumed to inform IDU proportions without bias (HANES, NDRI, and NSDUH) as can be seen in the decrease in deviance of the remaining data sources when one data source is removed (Table 7). Two remedies are available when data sources conflict: omit conflicting data sources, or explore different bias structures. We attempted to resolve the conflict by removing or assuming bias for the HANES IDU proportions, leaving only the CHS data to provide unbiased, non-mixed information, but this approach did not prove fruitful. The CHS data is not influential, with its removal from Model B5 not significantly changing deviance or quantity estimates (Table 7, row 3). In addition, ex-IDU proportions do not converge when the HANES IDU proportion data is removed or modeled with bias (Table 7, row 4). Removing the HANES or NSDUH data sources yields current IDU estimates higher than when any other data source is removed (Table 8, rows 5 and 9), largely driven by the NDRI estimate. Removing the NDRI data yields a severely low estimate for current IDU estimate (Table 8, row 7), much lower than that in the CHS data. Hence, removing any of these four data sources is not a viable solution.

We also considered scenarios where the NDRI or NSDUH data were modeled with bias. Given that overall prevalence is significantly lower when the NDRI or NSDUH data are removed (Table 8, rows 7 and 9), and we have a priori knowledge that the NDRI estimate of current IDUs, while biased slightly high, is more realistic compared to household surveys with likely underreported injecting drug use, we hesitated to remove the NDRI data. On the other hand, had we not treated either the HANES or CHS data as unbiased, the only other options for unbiased data on IDU proportions are data on high-risk subpopulations (JAILS, NHBS, HONE, STD) and data mixed across age groups (NDRI, NSDUH), both of which would likely increase the uncertainty of estimates.

Modeling the NDRI data with bias gives results similar to when it is omitted in the leave-one-data-source-out cross-validation framework (Tables 6 and 7, row 6). However, modeling the NSDUH data with bias returns an alternative to Model B5. With estimated biases of $e^\beta$ = 0.02 for current IDU proportion and $e^\beta$ = 0.5 for ex-IDUs, modeling the NSDUH data with bias reduces the estimated ex-IDU proportion to 2.32% from 2.53% in the model presented in the previous section, and less drastically, the current-IDU proportion from 0.58% to 0.56%. However, a drastic decrease in HCV prevalence among current and ex-IDUs, from 50.2% and 58.5% respectively to 44.7% and 30.5%, again reveals the sensitivity of these HCV prevalence estimates. This decrease, combined with slightly worse goodness-of-fit, leads us to prefer the model presented in Section 4.2. More unbiased data would allow us to further validate the modeling choices made for IDU proportions.

Table 7    Change in deviance of data sources using leave-one-data-source-out cross-validation

| Data Source Removed | CHC | CHS | HANES | HONE | NDRI | NHBS | NSDUH | JAILS | RISK | STD |
|---|---|---|---|---|---|---|---|---|---|---|
| None (Model B5) | 22 | 37 | 49,520 | 36 | 65,370 | 3,924 | 9,887 | 2.1 | 1,007 | 154 |
| CHC | - | 37 | 49,540 | 36 | 65,350 | 3,924 | 9,894 | 2.1 | 1,007 | 153 |
| CHS | 22 | - | 49,580 | 36 | 65,370 | 3,924 | 9,875 | 2.0 | 1,008 | 154 |
| HANES | * | * | - | * | **7.2** | 3,924 | **372** | * | * | * |
| HONE | 21 | 37 | 49,510 | - | 65,380 | 3,925 | 9,890 | 2.0 | 1,008 | 158 |
| NDRI | **15** | **19** | **12,630** | 40 | - | 3,924 | **11,262** | 2.1 | **1,358** | **256** |
| NHBS | 21 | 37 | 49,530 | 36 | 65,370 | - | 9,884 | 1.9 | **894** | 155 |
| NSDUH | **87** | **13** | **70** | 46 | **4.8** | 3,921 | - | 2.0 | 1,132 | **219** |
| JAILS | 21 | 37 | 49,530 | 36 | 65,370 | 3,925 | 9,882 | - | 1,008 | 154 |
| RISK | 22 | 37 | 49,450 | 37 | 65,540 | **68** | 9,779 | 2.0 | - | * |
| STD | 22 | 37 | 49,460 | 43 | 65,420 | 3,926 | 9,873 | 1.9 | 1,011 | - |

* indicates convergence not achieved according to the Gelman-Rubin diagnostic. All models besides Model B5 are based on 6,000 iterations per chain of 2 parallel chains. Model B5 uses 46,000 iterations per chain. The first 4,000 iterations of all chains were discarded as burn-in.

Table 8    Changes in parameter estimates using leave-one-data-source-out cross-validation

| Data Source Removed | $\rho_{cur}$ | $\rho_{ex}$ | $\pi_{cur}$ | $\pi_{ex}$ | $\pi_{non}$ | $\pi$ |
|---|---|---|---|---|---|---|
| All (prior model) | 23.8 (0 – 65.8) | 25.9 (0 – 69.2) | 49.9 (0 – 100) | 47.7 (0 – 98.3) | 50 (0 – 100) | 49.2 (0.2 – 99.6) |
| None (Model B5) | 0.58 (0.57 – 0.59) | 2.53 (2.51 – 2.54) | 50.2 (47.0 – 53.4) | 58.5 (52.3 – 64.5) | 1.04 (1.03 – 1.06) | 2.78 (2.61 – 2.94) |
| CHC | 0.58 (0.57 – 0.59) | 2.53 (2.51 – 2.54) | 50.2 (47.0 – 53.5) | 58.5 (52.3 – 64.5) | 1.04 (1.03 – 1.06) | 2.78 (2.62 – 2.94) |
| CHS | 0.58 (0.57 – 0.59) | 2.53 (2.51 – 2.54) | 50.2 (47.0 – 53.4) | 58.5 (52.1 – 64.7) | 1.04 (1.03 – 1.06) | 2.78 (2.61 – 2.94) |
| HANES | **1.65** (1.64 – 1.67) | * | **39.8** (34.9 – 44.8) | * | * | * |
| HONE | 0.58 (0.57 – 0.59) | 2.53 (2.51 – 2.54) | 50.2 (47.0 – 53.4) | 58.5 (52.2 – 64.7) | 1.04 (1.03 – 1.06) | 2.78 (2.61 – 2.94) |
| NDRI | **0.30** (0.29 – 0.30) | **1.96** (1.95 – 1.98) | **55.3** (52.3 – 58.3) | **62.9** (56.5 – 69.1) | 1.04 (1.02 – 1.05) | **2.41** (2.28 – 2.54) |
| NHBS | 0.58 (0.57 – 0.59) | 2.53 (2.51 – 2.54) | **55.3** (52.2 – 58.4) | 59.7 (53.7 – 65.5) | 1.04 (1.03 – 1.06) | **2.84** (2.68 – 3.00) |
| NSDUH | **1.66** (1.64 – 1.67) | **0.07** (0.05 – 0.10) | **66.4** (62.4 – 69.9) | 69.6 (64.4 – 74.5) | 1.04 (1.02 – 1.05) | **2.17** (2.10 – 2.24) |
| JAILS | 0.58 (0.57 – 0.59) | 2.53 (2.51 – 2.54) | 50.2 (47.1 – 53.4) | 58.3 (52.0 – 64.5) | 1.04 (1.03 – 1.06) | 2.77 (2.61 – 2.94) |
| RISK | 0.58 (0.57 – 0.58) | 2.53 (2.51 – 2.54) | * | * | 1.04 (1.03 – 1.06) | * |
| STD | 0.58 (0.57 – 0.59) | 2.53 (2.51 – 2.54) | 50.8 (47.6 – 54.1) | 58.3 (52.2 – 64.6) | 1.04 (1.03 – 1.06) | 2.78 (2.61 – 2.94) |

\* indicates convergence not achieved according to the Gelman-Rubin diagnostic. All models besides Model B5 are based on 6,000 iterations per chain of 2 parallel chains. Model B5 uses 46,000 iterations per chain. The first 4,000 iterations of all chains were discarded as burn-in.

# 6 Discussion

The estimates of IDU proportions and HCV prevalence depend on the model structure, including which data sources are assumed to be unbiased and assumed bias structures. We have investigated alternative models by varying which data sources are considered unbiased, assuming different bias strucures, and performing leave-one-data-source-out cross-validation. The model presented in Section 5 remains the leading model on which we base our discussion. However, we emphasize that more unbiased data are needed to further validate the model.

## 6.1 Comparison to Existing Estimates

As noted in the New York City Department of Health's Hepatitis C action plan, "it is likely that HCV infection prevalence in New York City lies somewhere in the range of 1.8 to 2.4%, though it could be as high as 3.8%" [NYCDOHMH 2013]. This study estimates that as of 2012, overall HCV prevalence of NYC adults aged 20-59 is 2.78% (95% CI 2.61% – 2.94%). Our estimate is higher than existing estimates. The 2004 HANES, a household survey designed to be representative of the NYC non-institutionalized adult population, estimated HCV prevalence at 2.2% (95% CI 1.5% to 3.3%) or 129,000 individuals [Bornschlegel 2009]. In NYC, health care providers and laboratories are required to report positive Hepatitis C tests to the Department of Health. Balter et al. adjusted surveillance data from 2000-2010 to account for biases arising from death, migration, under-diagnosis, etc., yielding an estimate of 2.37% (95% CI 1.5% to 4.9%) or 146,500 individuals [Balter 2014].

The HANES and adjusted surveillance estimates are for NYC adults aged 20 years old and above, while ours only covers adults aged 20-59. To compare our study to these estimates, we compute *absolute numbers* of infected individuals based on the size of the NYC 20-59 population. For HANES, we use an age-stratified version[4] of the 2.2% estimate described above from Bornschlegel et al. [Bornschlegel 2009]. Because the adjusted surveillance estimate [Balter 2014] is neither age nor IDU risk-group stratified, we use its overall estimate of 2.37%. For our study, we use age and risk-group stratified HCV prevalence estimates from Table 3. This results in the following number of individuals aged 20-59 estimated to be infected with HCV for our study, HANES, and adjusted surveillance respectively: 132,500 individuals (95% CI 124,900 to 140,000), 105,000 individuals (95% CI 71,500 to 157,500), and 113,100 individuals (95% CI 71,500 to 233,900).

We also compare estimates for young adults aged 20-39. HANES estimates 7,500 (95% CI 2,400 – 28,750) young NYC adults aged 20-39 to be infected with HCV, whereas our study obtains an estimate of 33,900 individuals (95% CI 29,300 – 38,800), which is 4.5 times more when comparing means[5]. We discuss the implications of these findings in the next section.

Our model also yields an updated estimate for the number of current IDU aged 20-59

---

[4] The HANES age-stratified estimates, not taking into account IDU risk group, are 0% (0-0%), 0.6% (0.2-2.3%), 2.5% (1.3-4.8%), and 5.8% (3.3-10%) for age groups 20-29, 30-39, 40-49, and 50-59 [Bornschlegel 2009]. Our model uses a version of these HANES age-stratified estimates for non-IDUs (0%, 0.4%, 1.8%, 2.4% for the same age categories).

[5] Though the upper limit of the 95% CI of the HANES estimate for young adults aged 20-39 is not far from our estimate's 95% CI's lower limit.

years, which is smaller than past estimates using multiplier methods [Brady 2008, Tempalski 2013]. We estimate that 27,600 (95% CI 27,300 – 27,800) individuals in NYC had injected drugs in the past year. There are not many estimates of NYC current injecting drug use. An estimate published in 2002 was for NYMSA, reporting that there were 143,000 current IDUs in the NYMSA [Brady 2008]. This estimate, when updated for 2007 [Tempalski 2013], had decreased by half. This 2007 estimate is one of our ten primary data sources; we adjusted it down to the NYC level from NYMSA as detailed in Section 3. Our geographic focus on NYC rather than the much larger NYMSA, as well as the contribution of two other data sources to inform the current IDU estimate could explain the difference between our model's estimate and the original 2007 estimate.

## 6.2   Implications

**HCV prevalence in NYC is higher than previously indicated from household surveys and surveillance systems.** The estimate of the total size of the HCV-positive population from our study is higher than the estimates from HANES and the surveillance study, but within their 95% confidence intervals. Subpopulations at greatest risk for HCV in NYC include IDUs, the incarcerated, and the homeless [Balter 2014]. Unlike the other studies, our study explicitly accounts for higher HCV prevalence among IDUs by stratifying the NYC population by IDU status. We also include data from the incarcerated population that was not included in HANES. While surveillance covers settings such as jails, drug treatment facilities, and needle exchange programs, it does not specify which of these risk factors affect which individuals in the data [NYCDOHMH 2013]. We explicitly account for each risk factor by using data sources specific to each setting and modeling data-source specific bias. In this way, we were able to include data sources covering non-representative subpopulations such as the incarcerated, communities of immigrants from hepatitis C endemic countries, patients at low-cost community health clinics, etc. These data sources are essential, contributing key information on current and ex-IDUs likely underrepresented in household surveys such as HANES. While we do not explicitly account for homelessness, the CHC data provides some information on this population, with 16% of respondents being homeless [CHC 2012].

Our analysis is the first to systematically include data from two NYC subpopulations with emerging risk of hepatitis C infection [NYCDOHMH 2013]: men who have sex with men who engaged in high-risk sexual activities (captured in the NHBS data) and young adults who have started to inject drugs.

**HCV transmission has increased among young injecting adults in NYC.** Recent studies have linked increased HCV transmission among young adults to the burgeoning opioid epidemic and associated injection drug use among young adults [CDC 2015; Zibbell 2017]. Our results show that among younger NYC adults (20-39 years old), injecting drug users have higher HCV prevalence than their non-IDU counterparts, which suggests that increasing HCV transmission among young injecting drug users in New York state [CDC 2008] and other parts of the US [Suryaprasad 2014; CDC 2011; CDC 2015] also applies in NYC.

**Baby boomers, regardless of injecting drug use status, carry a significant HCV burden.** Adults aged 40-59 (born 1953-1972) have the highest HCV prevalence of any age group, consistent with previous studies that attributed HCV transmission among this generation of individuals to unsafe medical procedures of that time and contaminated blood pre-widespread blood screening procedures [Spaulding 2016; Ward 2013].

**NYC HCV prevalence is higher than nationally.** Our results indicate that the chronic hepatitis C disease burden in NYC is higher than the national average, with national HCV prevalence estimated to be 1.3% from the 2002 NHANES [Armstrong 2006], the national survey on which HANES is based.

## 6.3 Limitations

There are several limitations to these results. First, because data for individuals older than 59 were not available in several of our data sources, our estimates do not include individuals older than 59. However, they do include the age group of 50-59 years, which has been suggested as the peak age group for HCV prevalence [Balter 2014] at present. Similar data constraints prevented us from including individuals under age 20. As noted by Balter et al., excluding individuals under age 20 makes our results more comparable with local and national estimates, and is unlikely to change the absolute total, as only a small proportion of this population (<1%) is estimated to be HCV positive [Balter 2014].

The leave-one-data-source-out cross-validation found that the model relies heavily on the RISK and HANES data sources, with estimates not converging when they are removed. Additional unbiased information on the quantities that RISK and HANES inform would reduce the dependence of the model on two data sources. In particular, unbiased information on IDU proportions not mixed across age groups should increase the precision of estimates for IDU proportions, and can also be used to confirm predictions made by the current model for these estimates. More data is also needed for current and ex-IDU prevalence to reduce the uncertainty around the estimation of these quantities.

The bias structure in the final model allows bias to vary by IDU risk group and data source. However, differential bias by age groups may be present and not accounted for in the alternative bias structures considered, which may be unreasonable if risk factors for injecting drug use or hepatitis C differ by age group. If data are available, further stratification on gender, country of origin, and any other variables that capture variation in IDU status or hepatitis C infection can improve this estimate.

The assumptions that the distributions of D, TSS, and AAFU are independent of each other and do not vary across age groups may be strong but are unavoidable for our data due to the lack of information – each of these variables have seven to ten categories to allow drug use history to be captured in detail due to its pivotal role in the model. Further stratification by age group results in cells that are extremely sparse, hence finer-grained data is needed.

## 7 Conclusions

Utilizing ten NYC data sources on IDU and HCV prevalence, our model estimates that as of 2012, overall chronic HCV prevalence in NYC adults aged 20-59 is 2.78% (95% CI 2.61% – 2.94%), or 132,500 affected individuals (95% CI 124,900 – 140,000 individuals). These estimates suggest that HCV prevalence in NYC is higher than previously indicated from household surveys and surveillance systems, and HCV transmission has increased among young injecting adults in NYC. Our method produced an ancillary benefit: an updated NYC-specific estimate of current IDU prevalence at 0.58% or 27,600 individuals.

Local prevalence estimates of HCV are increasingly important. With the advent of improved treatment regimes for hepatitis C such as Direct Acting Antiviral therapy (DAA), it may now be possible to not just prevent, but also treat hepatitis C in many individuals. NYC's urban setting with high-risk subpopulations means national estimates do not fully describe local trends in HCV transmission. Hence, accurate local prevalence estimates are important for local policymakers to better allocate resources for HCV prevention and treatment efforts, particularly in light of the emerging opioid epidemic and associated increase in hepatitis C infections.

# Acknowledgements

We thank the New York City Department of Health and Mental Hygiene, New York City Department of Correction, National Development and Research Institutes, Inc., Beth Israel Medical Center, and Mount Sinai Medical Center for providing data. We also thank Jay Varma, Tiffany Harris, Eric Rude, Ashly Jordan, Katherine Bornschlegel, Julie Schillinger, Jennifer Norton, Leena Gupta, Alan Neaigus, Kathleen Reilly, Farah Parvez, Howard Alper, Sam Friedman, Barbara Tempalski, Enrique Pouget, Brooke West, Don Des Jarlais, Kamyar Arasteh, Ponni Perumalswami, Holly Hagan, Janette Yung, Joseph Kennedy, and Eric Rude for helpful comments and/or data assistance at various stages of this project. We also thank two anonymous reviewers for their insightful comments.

# Declaration of Conflicting Interests

The authors declare that there is no conflict of interest.

# Appendix

### Drug Use History Information

The three drug use history variables, Injecting duration D, time since starting injecting TSS, and age at first use AAFU, are related: let $T_p$ be the present time (defined as 2012 in this study), $T_1$ be the time of first use, and $T_2$ be the time of last use. Then TSS = $T_p - T_1$, D = $T_2 - T_1$ for ex-IDUs, and AAFU = A – TSS where A represents age. D is undefined for current IDUs since their $T_2$ is some future unknown time.

### Cessation of injecting drug use

Sweeting et al. showed that $\kappa_{ex,a}$, the probability an injecting drug user has stopped injecting, can be written as:

$$\kappa_{ex,a} = \frac{\sum_{t=0}^{T} F_{D|ever}(t) f_{TSS|ever}(t) \left( F_{AAFU|ever}((a_{i+1})-t) - F_{AAFU|ever}(a_i - t) \right)}{\sum_{t=0}^{T} f_{TSS|ever}(t) \left( F_{AAFU|ever}((a_{i+1})-t) - F_{AAFU|ever}(a_i - t) \right)} \quad \text{(A1)}$$

where $F_{D|ever}(t)$ is the cumulative distribution function of D among ever-IDUs, $f_{TSS}$ is the probability distribution function of TSS, and $F_{aafu}(a)$ is the probability that AAFU is less than or equal to age group $a$ ($a_{i+1}$ denotes the age group following age group $a_i$, e.g., if $a_i$ = 30-39 years, then $a_{i+1}$ = 40-49 years). The simplifying assumption that distributions of drug use history variables D, TSS, and AAFU are independent of each other and invariant across age groups was made to obtain Equation (A1).

### Relating Drug Use History Information Between IDUs

Sweeting et al. propagated information on D and TSS across current, ex-, and ever-IDUs through the following relationships:

$$f_{D|ex,a}(t) = \frac{f_{D|ever,a}(t)[1-F_{TSS|ever,D=t,a}(t)]}{\kappa_{ex,a}} \quad (A2)$$

$$f_{TSS|cur,a}(t) = \frac{f_{TSS|ever,a}(t)[1-F_{D|ever,TSS=t,a}(t)]}{1-\kappa_{ex,a}} \quad (A3)$$

$$f_{TSS|ex,a}(t) = \frac{f_{TSS|ever,a}(t)F_{D|ever,TSS=t,a}(t)}{\kappa_{ex,a}} \quad (A4)$$

For proofs of Equations (A1-4) we refer the reader to Sweeting et al.'s supplementary material [Sweeting 2008].

### Proof of New Equations to Adjust $f_{aafu|cur}$ and $f_{aafu|ex}$ for $f_{aafu|ever}$

$$f_{AAFU|cur,a}(a-t) = P(A-TSS = a-t \,|\, T_1 \leq T, T_2 \geq T, A=a)$$
$$= \frac{P(A-TSS=a-t\,|\,T_1\leq T,A=a)P(T_2\geq T\,|\,T_1\leq T,A-TSS=a-t,A=a)}{P(T_2\geq T\,|\,T_1\leq T,A=a)}$$
$$= \frac{f_{AAFU|ever,a}(a-t)P(T_2-T_1\geq T-T_1\,|\,T_1\leq T,A-TSS=a-t,A=a)}{1-\kappa_{ex,a}}$$
$$= \frac{f_{AAFU|ever,a}(a-t)[1-F_{D|ever,TSS=t,a}(t)]}{1-\kappa_{ex,a}} \quad (A5)$$

$$f_{AAFU|ex,a}(a-t) = P(A-TSS = a-t \,|\, T_1 < T, T_2 < T, A=a)$$
$$= \frac{P(A-TSS=a-t\,|\,T_1<T,A=a)P(T_2<T\,|\,T_1<T,A-TSS=a-t,A=a)}{P(T_2<T\,|\,T_1<T,A=a)}$$
$$= \frac{f_{AAFU|ever,a}(a-t)P(T_2-T_1<T-T_1\,|\,T_1<T,A-TSS=a-t,A=a)}{\kappa_{ex,a}}$$
$$= \frac{f_{AAFU|ever,a}(a-t)F_{D|ever,TSS=t,a}(t)}{\kappa_{ex,a}} \quad (A6)$$